\documentclass{iopconfser}

\usepackage{amsmath,amssymb,amsfonts,latexsym,cancel}
\newcommand{\be}{\begin{equation}}
\newcommand{\ee}{\end{equation}}
\newcommand{\bea}{\begin{eqnarray}}
\newcommand{\eea}{\end{eqnarray}}

\begin{document}

\title{Photon helicity asymmetry and duality anomaly in rotating waveguides}

\author{Adrian del Rio$^1$} 
\affil{\it $^1$Universidad Carlos III de Madrid, Departamento de Matem\'aticas. \\ Avenida de la Universidad 30 (edificio Sabatini), 28911 Legan\'es (Madrid), Spain.}

\email{adrdelri@math.uc3m.es}

\begin{abstract}
In quantum field theory, photon pairs can be  excited out of the  vacuum under the influence of dynamical gravitational fields. If  the classical electromagnetic duality symmetry held at the quantum level, equal numbers of right- vs left-handed modes would be excited. However, a quantum chiral-like anomaly is known to affect this symmetry. In this work, we explore more experimentally accessible manifestations of this  anomaly in analogue-gravity systems. Specifically, we describe the physical mechanism by which rotating, accelerating waveguides can induce an imbalance between right- and left-handed photons excited from the  vacuum---effectively signaling the emergence of the electromagnetic duality anomaly in controlled laboratory settings. To build physical intuition, we review first the Adler-Bell-Jackiw anomaly in 1+1 dimensions, which elegantly illustrates  the connection between asymmetric particle pair creation and chiral anomalies.

\end{abstract}


\section*{Motivation: electric-magnetic duality and anomaly in gravity.}

Let $F_{ab}$  and ${^*F}_{ab}$ denote the electromagnetic field tensor and  its Hodge dual, respectively, on a 3+1 dimensional globally hyperbolic spacetime, with Levi-Civita connection $\nabla_a$. Maxwell's equations in vacuum, $\nabla_a F^{ab}=\nabla_a {^*F}^{ab}=0$,  which govern the propagation of free electromagnetic waves,  exhibit an invariance under duality rotations that mix the electric and magnetic components as $F_{ab}\to \cos\theta F_{ab} +\sin\theta\,  {^*F}_{ab}$, for $\theta\in \mathbb R$. This transformation leads to a Noether symmetry of the classical source-free action,  even in curved spacetimes \cite{PhysRevD.13.1592},  generated in phase space by the following conserved Noether charge
\bea
Q(A) = \frac{1}{2} \int_\Sigma d\Sigma_b \left( A_a {^*F}^{ba} - {Z}_a F^{ba} \right)\, , \label{Q}
\eea
where $\Sigma$ is an arbitrary Cauchy surface, $A_a$ is the magnetic potential (with $F=dA$), and $Z_a$ is the dual ``electric'' potential  (with ${^*F}=dZ$,  only available in the absence of charges). Physically, (\ref{Q})  measures the net difference between right- and left-handed circularly polarized electromagnetic modes  \cite{10.1119/1.1971089}, corresponding to the Stokes V parameter in optics. 
As a result, the circular polarization state of classical electromagnetic waves is conserved during propagation, even in the presence of strong gravitational fields. 

In contrast, when the electromagnetic field is quantized on a generic curved spacetime, the vacuum expectation value of the quantum counterpart of (\ref{Q}) is no longer conserved in time, with its time-dependence determined by the spacetime geometry \cite{PhysRevLett.118.111301, PhysRevD.98.125001, doi:10.1142/S0218271817420019, sym10120763}. For asymptotically flat spacetimes, this quantum anomaly \cite{10.1093/acprof:oso/9780198507628.001.0001}  was shown to arise if the gravitational background carries a flux of circularly polarized gravitational waves \cite{PhysRevLett.124.211301, PhysRevD.104.065012, PhysRevD.108.044052}, or equivalently, if the spacetime has some degree of helicity.

Now, in realistic astrophysical scenarios, the observable consequences of this  quantum effect are expected to be extremely small. This motivates the search for other, more accessible manifestations of the electromagnetic duality anomaly --- in particular, in analogue-gravity systems, which are rapidly developing as platforms to simulate relativistic quantum effects in a controlled laboratory setting \cite{Barcel__2005}.  The main goal of this work is to present and analyze one such example \cite{anomaly}.

 \section*{A simple model: Adler-Bell-Jackiw anomaly in 1+1 dimensions.}
 
 To gain intuition, it is useful to consider first a simpler model, namely the Adler-Bell-Jackiw anomaly in 1+1 dimensions \cite{1985caa..book..211J}. This is the simplest example of a chiral anomaly,  and illustrates one of the key points of this work: the connection between chiral anomalies and asymmetric particle pair creation.

 Let us consider a massless  Dirac field $\Psi$ in a flat spacetime ($\mathbb R\times \mathbb S^1$, $\eta_{ab}$),  coupled to an external electromagnetic field $F_{ab}$, with magnetic potential $A_a$. The Dirac  equation reads
 $
 i\gamma^a(\nabla_a-i q A_a)\Psi=0\, ,
 $
 where $q$ is the fermion's charge, and $\gamma^a$ are the Dirac  matrices satisfying the Clifford algebra, $\{\gamma^a, \gamma^b\}=2\eta^{ab}\mathbb I_{2\times 2}$. In 1+1 dimensions this algebra is generated by only two  matrices, and  the commutator $[\gamma^a,\gamma^b]=2\epsilon^{ab}\gamma_2$ ---with $\epsilon^{ab}$  the totally antisymmetric tensor--- defines the chiral matrix $\gamma_2$, yielding $\gamma_2=\gamma^0\gamma^1$.  
 
  The eigenvectors of the chiral matrix,  satisfying $\gamma_2u_{\pm}=\pm u_{\pm}$, can be used to decompose the spinor field  in  two  chiral components: $\Psi=u_+ + u_-$.
 In 1+1 dimensions,  chirality physically represents the direction of propagation. Namely, the ``right-handed'' spinor $u_+$ propagates to the right, while the ``left-handed'' one is a left-moving field. 
 Interestingly, the Dirac  action is invariant under chiral transformations
 \bea
 \Psi \to  e^{i\gamma_2 \theta}\Psi=e^{i \theta} u_+ + e^{-i \theta} u_-\, ,\quad \theta\in \mathbb R\, ,  \label{chiralT}
 \eea
 which rotate the two chiral components in opposite directions. This symmetry is generated by a conserved  Noether charge,  $Q_2=||u_+||^2-||u_-||^2$, written in terms of the Dirac inner product.  However, at the quantum level, and external electric field $E=\epsilon^{ab}F_{ab}$ spoils this conservation in time:
 \bea
 \langle {\rm in}| \hat Q_2(t_1) |{\rm in} \rangle - \langle  {\rm in}|  \hat Q_2(t_0)|{\rm in} \rangle = 2\hbar \int_{t_0}^{t_1} dt\, E(t,x)\neq 0 \, .\label{ABJ}
 \eea

The existence of this anomaly can be understood in terms of particle pair creation  \cite{PhysRevD.108.105025}. For simplicity, we consider a spatially homogeneous  field, $F_{ab}=F_{ab}(t)$. If the electric field is non-zero only during a finite time interval $(t_0,t_1)$, we can define natural ``in'' and ``out'' notions of vacua and particles, associated with early ($t\leq t_0$) and late times $(t\geq t_1)$. Now, even though the external  electric field may vanish at late times, $F_{ab}=0$, the quantum system does not really return to its original configuration, because a residual gauge potential $A_a\neq 0$ can remain after the pulse. For example, in the gauge $A_t=0$, we can have $A_a=(0,A)$ with $A=\int_{t_0}^{t_1}E(t)dt$, provided $A_a=0$ at early times  (assume $qA>0$, without loss of generality). While this residual gauge potential has no classical effect, it  does induce a {\it permanent shift} in the frequency spectrum of the fermion modes,  as a result of which the ``in'' and ``out'' basis can differ. Specifically, if the ``in'' modes that solve the Dirac  equation at early times are given by 
\bea
u^{\text{in}}_{+,n}(t,x) \sim e^{-i \omega_n(t - x)} \begin{pmatrix} 1\\ 0 \end{pmatrix}\, ,  \quad u^{\text{in}}_{-,n}(t,x) \sim e^{-i \omega_n(t + x)} \begin{pmatrix} 0\\ 1 \end{pmatrix} \, ,
\eea
with oscillatation frequency  $\omega_n=n\in \mathbb Z$,  then the ``out'' modes that solve the field equations at late times
\bea
u^{\text{out}}_{+,n}(t,x) \sim e^{-i n(t - x)} e^{- i q A t} \begin{pmatrix} 1\\ 0 \end{pmatrix}\, ,  \quad u^{\text{out}}_{-,n}(t,x) \sim e^{-i n(t + x)} e^{+ i q A t} \begin{pmatrix} 0\\1 \end{pmatrix}\, , \label{lategauge}
\eea
 oscillate with frequency $\omega^{\pm}_n=n\pm qA$. This is, the out modes are simply a (time-dependent) chiral transformation  of the in modes, of the form (\ref{chiralT}) with parameter $\theta(t)\equiv q A t$. 
 
 This (asymmetric) shift in frequency implies that a finite number of modes that had positive frequency at early times now acquire negative frequency at late times, or viceversa.  Most importantly, the asymmetry between the two chiral sectors induces a net transfer of modes from negative to positive helicity. To see this, let us write the ``in'' representation of the quantum chiral fields  as
 \bea
\hat u_\pm(t,x) &=& \sum_{n>0}^{\infty} a^{\text{in}}_{\pm,n} u^{\text{in}}_{\pm,n}(t,x) + \sum_{n>0}^{\infty} b^{\text{in}\dagger}_{\mp,n} u^{\text{in}}_{\pm,-n}(t,x)\label{1} \, .
  \eea
 For massless fields, helicity agrees with chirality for particles, and differ in sign for anti-particles. Thus, in the right-handed sector (equation $+$ in (\ref{1})), the positive-frequency modes $\{u^{\text{in}}_{+,n}\}_{n \in \mathbb N}$ have positive helicity $+\hbar$, while the negative-frequency modes $\{u^{\text{in}}_{+,-n}\}_{n\in \mathbb N}$ have negative helicity $-\hbar$. Similarly, in the left-handed sector (equation $-$ in (\ref{1})),  the positive-frequency modes $\{u^{\text{in}}_{-,n}\}_{n\in \mathbb N}$ have negative helicity $-\hbar$, while  $\{u^{\text{in}}_{-,-n}\}_{n\in \mathbb N}$ have positive helicity $+\hbar$. Now, the ``out'' representation reads 
  \bea
 \hat u_+ = \sum_{n>-qA}^{\infty} a^{\text{out}}_{+,n} u^{\text{out}}_{+,n} + \sum_{n>qA}^{\infty} b^{\text{out}\dagger}_{-,n} u^{\text{out}}_{+,-n} \, , \quad\quad
\hat u_- = \sum_{n>qA}^{\infty} a^{\text{out}}_{-,n} u^{\text{out}}_{-,n} + \sum_{n>-qA}^{\infty} b^{\text{out}\dagger}_{+,n} u^{\text{out}}_{-,-n} \label{4}\, ,
 \eea
 which contains a different number of helicity modes in each chiral sector as compared to the in basis. More precisely, some modes originally contained in the second sum of $u_+$ in (\ref{4}) have now moved to the first sum, thus flipping their helicity from negative to positive. Similarly,  some modes originally contained in the first sume of $u_-$ in (\ref{4}), have now moved to the second sum, flipping their helicity value also from negative to positive. By normal-ordering the operator $\hat Q_2=||\hat u_+||^2-||\hat u_-||^2$ using the mode expansions in (\ref{4}), and using  Bogoliubov transformations relating the ``in'' and ``out'' vacua,  one is able to find
\bea
\langle {\rm in}| :Q_2:(t_{\rm 1}) |{\rm in}\rangle - \langle {\rm in}| :Q_2:(t_{\rm 0}) |{\rm in}\rangle &=& \hbar \left[ \sum_{n>-qA}^\infty 1 - \sum_{n>-qA}^\infty 1 \right] = 2\hbar [qA]\, , \label{finalgauge}
\eea
which exactly matches the prediction from the chiral anomaly (here $[\cdot]$ indicates integer part;  a vacuum polarization term is missing in normal-ordering, see \cite{PhysRevD.108.105025} for more details).
In conclusion, asymmetric particle creation is the physical mechanism behind the emergence of the chiral anomaly.

 \section*{Electromagnetic duality anomaly in a 3+1-dimensional accelerating waveguide.} 

The question we want to explore now is whether a similar duality anomaly can arise in 3+1 dimensions without gravity, under the right conditions. As argued above,   chiral anomalies manifest when the background has some degree of helicity. Because of this, we consider an infinitely long,  empty cylindrical waveguide in flat Minkowski space of radius $R$. The waveguide is initially static, and then undergoes a period of acceleration -- both tangentially and longitudinally -- until it reaches a constant angular velocity $\Omega$ and a constant linear velocity $v$ along its symmetry axis. For simplicity, we restrict here to $\Omega>0$.

To obtain a quantum description of the electromagnetic field inside the waveguide we first need to specify the classical theory. We consider the following covariant phase space \cite{10.1063/1.528801, PhysRevD.103.025011}: 
\bea
\Gamma=\left\{A \in \Lambda^{1}(\mathbb{R}^{4})\, /\, \nabla_{a} F^{a b}=0,\,  \nabla_a A^a=0 \, , B(A)=0 \right\}\, .
\eea
The precise  form of the boundary conditions $B(A)$ depends on the physical  nature of the waveguide, effectively modeling how the electromagnetic modes interact with the microscopic constituents of this background. Depending on $B(A)$,  duality transformations may fail to be a symmetry already at the classical level. To ensure the classical symmetry,  we  have chosen specific boundary conditions that also remain invariant under  duality rotations \cite{anomaly}. In this model, $\Gamma$  can be endowed with a canonical symplectic structure $\Omega$ free from boundary contributions, which agrees with the usual one in free spacetime. This geometric structure allows us to  identify  the conserved Noether charge that generates duality rotations. The result, in this model, gives again equation (\ref{Q}). 

Now, to make the analogy with the fermionic system more direct, it is convenient to introduce right- and left-handed potentials, $A_a^R$ and $A^L_a$. These are defined so that their associated field strengths are self-dual and anti-self-dual, respectively: $i{^*F}^{\pm}_{ab}=\pm F^{\pm}_{ab}$. In terms of these complex variables, duality rotations act as simple chiral transformations: $F_{ab}^\pm \to e^{\mp i \theta }F_{ab}^{\pm }$.

Quantization  can be carried out using standard tools in QFT \cite{Wald:1995yp}. The construction of a Fock space of photon states requires specification of a well-defined vacuum state. To achieve that, we introduce a complex structure on $\Gamma$, compatible with the symplectic structure $\Omega$, to distinguish positive- and negative-frequency modes \cite{Ashtekar1975zn}. Since the system is stationary at both early and late times, we have two natural possibilities. At early times, this frequency  decomposition is given by
\bea
\hat A_{a}^{R}=\sum_{n=1}^{\infty} \sum_{m=-\infty}^{\infty} \int_{-\infty}^{\infty} d k\left[a_{k m n}^{R, \mathrm{in}} A_{a, 1 k m n}^{R, \mathrm{in}}+{a_{k m n}^{L, \mathrm{in},\dagger}} \overline{A_{a, 1 k m n}^{L, \mathrm{in}}}\right]\, ,
\eea
where, in terms of a cylindrical  Newman-Penrose null tetrad $\{e^I_a\}_{I=0}^3=\{\mathbf {n_a}, \mathbf{\boldsymbol \ell_a}, \mathbf{m_a}, \mathbf{\bar m_a} \}$ \cite{Torres2003}, 
\bea
A_{a, h k m n}^{R, \mathrm{in}}&\sim&  e^{-i\left(h \omega_{k m n} {t}+k {z}+m {\phi}\right)}\left[\frac{j_{m n}}{R} J_{m+1}\left(\frac{j_{m n}}{R} {\rho}\right) \overline{{\mathbf{m}}_{\mathbf{a}}}-i\left(h \omega_{k m n}+k\right) {\boldsymbol{\ell}}_{\mathbf a} J_{m}\left(\frac{j_{m n}}{R} {\rho}\right)\right]\in \Gamma\, ,\label{n1}\\
A_{a, h k m n}^{L, \mathrm{in}}&\sim & e^{-i\left(h \omega_{k m n} {t}+k {z}+m {\phi}\right)}\left[\frac{j_{m n}}{R} J_{m-1}\left(\frac{j_{m n}}{R} {\rho}\right) {\mathbf{m}}_{\mathbf{a}}+i\left(h \omega_{k m n}+k\right) {\boldsymbol\ell}_{\mathbf{a}} J_{m}\left(\frac{j_{m n}}{R} {\rho}\right)\right]\in \Gamma\, , \label{n2}
\eea
with $j_{mn}$  the $n$th zero of the Bessel function $J_m$, and $h\in \{-1,+1\}$. These modes have well-defined frequency $\omega_{kmn}$, linear momentum $k$, and angular momentum $m$ along the symmetry axis of the waveguide. The  basis vectors are  parallel transported along the integral curves of  $u=\partial/\partial t$ at early times, i.e. they remain constant as seen early inertial observers of 4-velocity $u^a$: $u^a\nabla_a e^I_b=0$. The  physical degrees of freedom are obtained by projecting with $\{\bold{m_{a}}, \bold{\bar m_{a}}\}$, which carry the two transverse photon polarizations.

Since the  modes (\ref{n1})-(\ref{n2}) have well-definite spin-weight,  once the waveguide reaches uniform rotation at late times, they acquire a time-dependent phase factor due to frame-dragging, which rotate the left and right-handed modes  in opposite directions  (just like an ordinary chiral transformation). Specifically, one has $\bold{m_{a}}=e^{i\Omega_0 \gamma t}\bold{m^{0}_a}$,  where $\gamma^2=(1-v^2)^{-1}$ and  $\bold{m^{0}_a}$  is now parallel transported along the integral curves of the accelerating observers $u^a$ {\it at late times}: $u^a\nabla_a \bold{m^{0}_b}=0$.  This leads to the ``out'' representation, 
\bea
\hat A_{a}^{R}=\sum_{(h k m n) \in H_{h}^{+}} a_{h k m n}^{R, \text{out}} A_{a, h k m n}^{R, \text {out}}+\sum_{(h k m n) \in H_{h}^{-}} {a_{h k m n}^{L, \text {out},\dagger}} \overline{A_{a, h k m n}^{L, \text {out}}}\, , \label{outrep}
\eea
where the field modes at late times acquire the form
\bea
\bold{m^{0,a}}A_{a, h k m n}^{R, \text{out}} &\sim& e^{-i\left[\left(h \omega_{k m n}+m \Omega-k v\right) \gamma t+\left(k-h \omega_{k m n} v-m v \Omega\right) \gamma z+m \phi\right]}   \mathbf{e}^{-\mathbf{i} \mathbf{\Omega} \gamma \mathbf{t}}\, ,\\
\bold{\bar m^{0, a}}A_{a, h k m n}^{L, \text{out}} &\sim& e^{-i\left[\left(h \omega_{k m n}+m \Omega-k v\right) \gamma t+\left(k-h \omega_{k m n} v-m v \Omega\right) \gamma z+m \phi\right]}  \mathbf{e}^{+\mathbf{i} \mathbf{\Omega} \gamma \mathbf{t}}\, .
\eea
This frame-dragging effect is crucial. Just like in the 1+1 fermion model  discussed earlier (i.e. equation (\ref{lategauge})), this residual angular momentum produces an asymmetric shift in the frequency spectrum. In particular, the domain of positive-frequency modes in the right-handed sector is different from that of the left-handed one,
$
H_{h}^{ \pm }=\left\{(k, m, n) \in \mathbb{R} \times \mathbb{Z} \times \mathbb{N} / h \omega_{kmn}-m \Omega-k v \pm \Omega > 0\right\} 
$, and a {\it spectral asymmetry} arises at  late times. This is the key ingredient that  ultimately gives rise to an imbalance in photon pair creation. By normal-ordering the operator  (\ref{Q}) using (\ref{outrep}), and using Bogoliubov transformations relating the ``in'' and ``out'' mode expansions, a lengthy calculation gives \cite{anomaly} 
\bea
\langle {\rm in}|\hat   Q |_{t\to+\infty} | {\rm in}\rangle-\langle {\rm in}|\hat  Q |_{t\to-\infty} | {\rm in}\rangle &=&2\hbar \, R\, \left[ \sum_{(kmn)\in H^{-}_1 }1 -   \sum_{(kmn)\in H^{+}_1 }1 \right]\neq 0\, , \label{eq2}
\eea
which quantifies the difference  number of photons with positive and negative helicity created from the vacuum. This is a sum over photon modes that have flipped sign in frequency due to the time-dependent background, and is analogous to (\ref{finalgauge}). The sum is convergent, and nonzero  if both $\Omega, v\neq 0$.

\section*{Acknowledgements.}
I acknowledge support from the {\it Atracci\'on de Talento --- C\'esar Nombela} program, grant No. 2023-T1/TEC-29023, funded by Comunidad de Madrid (Spain), as well as from the Spanish National Grant PID2023-149560NB-C21, funded by MCIU/AEI/10.13039/501100011033/FEDER, UE.

\bibliography{iopart-num}

\providecommand{\newblock}{}
\begin{thebibliography}{10}
\expandafter\ifx\csname url\endcsname\relax
  \def\url#1{{\tt #1}}\fi
\expandafter\ifx\csname urlprefix\endcsname\relax\def\urlprefix{URL }\fi
\providecommand{\eprint}[2][]{\url{#2}}

\bibitem{PhysRevD.13.1592}
Deser S and Teitelboim C 1976 {\em Phys. Rev. D\/} {\bf 13}(6) 1592--1597

\bibitem{10.1119/1.1971089}
Calkin M~G 1965 {\em Am. J. Phys.\/} {\bf 33} 958--960

\bibitem{PhysRevLett.118.111301}
Agullo I, del Rio A and Navarro-Salas J 2017 {\em Phys. Rev. Lett.\/} {\bf
  118}(11) 111301

\bibitem{PhysRevD.98.125001}
Agullo I, del Rio A and Navarro-Salas J 2018 {\em Phys. Rev. D\/} {\bf 98}(12)
  125001

\bibitem{doi:10.1142/S0218271817420019}
Agullo I, del Rio A and Navarro-Salas J 2017 {\em Int. J. Mod. Phys. D\/} {\bf
  26} 1742001

\bibitem{sym10120763}
Agullo I, del Rio A and Navarro-Salas J 2018 {\em Symmetry\/} {\bf 10} ISSN
  2073-8994

\bibitem{10.1093/acprof:oso/9780198507628.001.0001}
Bertlmann R~A 2000 {\em Anomalies in Quantum Field Theory\/} (Oxford University
  Press)

\bibitem{PhysRevLett.124.211301}
del Rio A, Sanchis-Gual N, Mewes V, Agullo I, Font J~A and Navarro-Salas J 2020
  {\em Phys. Rev. Lett.\/} {\bf 124}(21) 211301

\bibitem{PhysRevD.104.065012}
del Rio A 2021 {\em Phys. Rev. D\/} {\bf 104}(6) 065012

\bibitem{PhysRevD.108.044052}
Sanchis-Gual N and del Rio A 2023 {\em Phys. Rev. D\/} {\bf 108}(4) 044052

\bibitem{Barcel__2005}
Barcelo C, Liberati S and Visser M 2011 {\em Living Rev. Relativ.\/} {\bf 14}

\bibitem{anomaly}
del Rio A 2025 (\textit{Preprint} \eprint{2506.18610})

\bibitem{1985caa..book..211J}
{Jackiw} R 1985 {\em Current Algebra and Anomalies. Edited by Jackiw R et al\/}
  (World Scientific Publishing) pp 211--359

\bibitem{PhysRevD.108.105025}
del R\'{\i}o A and Agullo I 2023 {\em Phys. Rev. D\/} {\bf 108}(10) 105025

\bibitem{10.1063/1.528801}
Lee J and Wald R~M 1990 {\em J. Math. Phys.\/} {\bf 31} 725--743

\bibitem{PhysRevD.103.025011}
Margalef-Bentabol J and Villase\~nor E~J~S 2021 {\em Phys. Rev. D\/} {\bf
  103}(2) 025011

\bibitem{Wald:1995yp}
Wald R~M 1995 {\em {Quantum Field Theory in Curved Space-Time and Black Hole
  Thermodynamics}\/} (Chicago, IL: University of Chicago Press)

\bibitem{Ashtekar1975zn}
Ashtekar A and Magnon A 1975 {\em Proc. Roy. Soc. Lond. A\/} {\bf 346} 375--394

\bibitem{Torres2003}
Torres~del Castillo G~M 2003 {\em {3D spinors, Spin-Weighted Functions and
  their Applications}\/} (Boston, USA: Birkhauser)

\end{thebibliography}

\end{document}